\documentclass[aps,12pt,final,notitlepage,oneside,onecolumn,showpacs,amsmath,amssymb]{revtex4-2}
\usepackage{color}

\begin{document}
\title{Anisotropic solutions for $R^2$ gravity model with a scalar field}
\author{\firstname{Vsevolod R.}~\surname{Ivanov}}
\email{vsvd.ivanov@gmail.com}
\affiliation{Physics Department, Lomonosov Moscow State University,\\
Leninskie Gory~1, 119991 Moscow, Russia}
\author{\firstname{Sergey~Yu.}~\surname{Vernov}}
\email{svernov@theory.sinp.msu.ru}
\affiliation{Skobeltsyn Institute of Nuclear Physics, Lomonosov Moscow State University,\\
 Leninskie Gory 1, 119991, Moscow, Russia}

\begin{abstract}
We study anisotropic solutions for the pure $R^2$ gravity model with a scalar field in the Bianchi I metric. The evolution equations have a singularity at zero value of the Ricci scalar $R$ for anisotropic solutions, whereas  these equations are smooth for isotropic solutions. So, there is no anisotropic solution with the Ricci scalar smoothly changing its sign during evolution. We have found anisotropic solutions using the conformal transformation of the metric and the Einstein frame. The general solution in  the Einstein frame has been found explicitly. The corresponding solution in the Jordan frame has been constructed in quadratures.
\end{abstract}



\maketitle

\section{Introduction}

The observable Universe is homogenous and isotropic at large scale and there are strong limits on
anisotropic models from observations~\cite{Bernui:2005pz,Appleby:2012as}. For that reason, the Friedmann--Lema\^{i}tre--Robertson--Walker (FLRW) metric plays the central role in the description of the global evolution of the Universe.
Models with scalar fields are actively investigated, because they can describe the observable evolution of the Universe as the dynamics of FLRW background and cosmological perturbations.

The mechanism of isotropization of anisotropic solutions is an important question.
The Wald theorem~\cite{Wald:1983ky} proves that all initially expanding Bianchi models except type IX approach the
de Sitter space-time if the energy conditions are satisfied. For space-time of Bianchi types I--VIII
with a positive cosmological constant and matter satisfying the
dominant and strong energy conditions, solutions which exist globally
in the future have certain asymptotic properties at
$t\rightarrow\infty$. Anisotropic solutions are actively studied both in general relativity models with minimally coupled scalar fields~\cite{Aguirregabiria:1993pm,Pereira:2007yy,Arefeva:2009hyz,Arefeva:2009tkq,Fadragas:2013ina,Paliathanasis:2021ztt,Chataignier:2022yic}, and
in modified gravity models, in particular, in models with nonminimally coupled scalar field~\cite{Starobinsky:1981st,Sami:2012uh,Kamenshchik:2016atu,Camci:2016yed,Kamenshchik:2017ous,Kamenshchik:2017ojc}, in $F(R)$ gravity models~\cite{Gurovich:1979st,Figueiro:2009mm,Leon:2010pu,Muller:2017nxg,Arora:2022dti,Nojiri:2022idp}, in models with the Ricci tensor squared term~\cite{Muller:2006xy,Barrow:2009gx,Toporensky:2016kss}, in the recently proposed~\cite{Ketov:2022lhx,Ketov:2022zhp} the Starobinsky--Bel--Robinson gravity model~\cite{Do:2023yvg}, and within the context of the Horndeski theory~\cite{Starobinsky:2019xdp,Galeev:2021xit}.

In this paper, we obtain the general solution in the Bianchi I metric for the pure $R^2$ model with a scalar field. By the conformal transformation of the metric, this model can be transformed to a two-field model with a nonstandard kinetic part, so-called chiral cosmological model~\cite{Chervon:1995jx,DiMarco:2002eb,Paliathanasis:2018vru,Chervon:2019nwq,Braglia:2020eai,Anguelova:2020nzl,Zhuravlev:2020ugb,Paliathanasis:2020sfe,Ivanov:2021ily,Diaz-Barron:2021ari,Tot:2022dpr}. Note that the metric transformation is well-defined only if the Ricci scalar $R$ does not change its sign during evolution. By this reason, some FLRW solutions have no analogues in the Einstein frame~\cite{Ivanov:2021ily}. By considering the evolution equations in the Bianchi I metric, we show that anisotropic solutions cannot smoothly pass the boundary $R=0$. So, we can use the Einstein frame to seek anisotropic solutions. We find solutions in the cosmic time for the considered two-field chiral cosmological model and use the inverse conformal transformation to get solutions for the initial $R^2$ model with a scalar field.

\section{$R^2$ model with a scalar field}
Let us consider a pure $R^2$ model, describing by the following action:
\begin{equation}\label{actR2}
    S_{R}=  \int {d}^4 x \sqrt{-\tilde{g}} \left[F_0\tilde{R}^2 - \frac{\varepsilon_\psi}{2} \tilde{g}^{\mu \nu} \nabla_{\mu}\psi \nabla_{\nu}\psi \right],
\end{equation}
where $F_0$ is a positive constant,  $\tilde{R}$ is the Ricci scalar,  $\psi$ is a scalar field or a phantom scalar field in dependence of the sign of  $\varepsilon_\psi=\pm1$.

The general solution in the case of the spatially flat Friedmann--Lema\^{i}tre--Robertson--Walker metric has been found in~\cite{Ivanov:2021ily}.
In this paper, we consider the case of the Bianchi I metric with the following interval~\cite{Gurovich:1979st,Pereira:2007yy}:
\begin{equation}
\label{BianchiIN}
 ds^2={}-d{\tilde{t}\,}^2+\tilde{a}^2(\tilde{t})\left[\mathrm{e}^{2\tilde{\beta}_1(\tilde{t})}\,dx_1^2+\mathrm{e}^{2\tilde{\beta}_2(\tilde{t})}\,dx_2^2+\mathrm{e}^{2\tilde{\beta}_3(\tilde{t})}\,dx_3^2\right].
\end{equation}

The functions $\tilde{\beta}_i(\tilde{t})$ satisfy the relation
\begin{equation}
\label{bianchi_constraint}
\tilde{\beta}_1(\tilde{t}) +  \tilde{\beta}_2(\tilde{t}) +  \tilde{\beta}_3(\tilde{t}) = 0.
\end{equation}

It is useful to introduce a shear,
\begin{equation}
 \tilde{\sigma}^2\equiv\dot{\tilde{\beta}}_1^2+\dot{\tilde{\beta}}_2^2+\dot{\tilde{\beta}}_3^2 = 2\left(\dot{\tilde{\beta}}_1^2 + \dot{\tilde{\beta}}_1\dot{\tilde{\beta}}_2+ \dot{\tilde{\beta}}_2^2\right)\,,
 \end{equation}
that measures a total amount of anisotropy. In this section, ``dots'' denote derivatives with respect to time $\tilde{t}$.

As known, the $F(R)$ gravity model has the following evolution equations:
\begin{equation}
\label{frequ}
    F_{,\tilde{R}}\tilde{R}_{\mu\nu}-\frac{1}{2}g_{\mu\nu}F-(\nabla_{\mu}\nabla_\nu-g_{\mu\nu}\Box)F_{,\tilde{R}}=\frac12 T_{\mu\nu},
\end{equation}
where $F_{,\tilde{R}}\equiv\frac{dF}{d\tilde{R}}$, $T_{\mu\nu}$ is the matter stress-energy tensor.

In the Bianchi I metric, the Ricci scalar is
\begin{equation}\label{Rs}
     \tilde{R}=\tilde{\sigma}^2+\tilde{R}_i,
\end{equation}
where
\begin{equation*}
    \tilde{R}_i=6\left(\dot{H}_J+2H_J^2\right),\qquad H_J=\frac{\dot{\tilde{a}}}{\tilde{a}}\,.
\end{equation*}

For $F(R)=F_0R^2$, system (\ref{frequ}) in the Bianchi I metric contains the following equations:
\begin{equation}
\label{equFR00}
\begin{aligned}
3 H_J \dot{\tilde{\sigma}}^2 - \frac34 \left({\tilde{\sigma}}^2\right)^2 - 3\left(2\dot{H}_J + 3 H_J^2\right) \tilde{\sigma}^2 + 18 H_J \ddot{H}_J - 9 \dot{H}_J^2 + 54 H_J^2 \dot{H}_J = \frac{\varepsilon_\psi}{8 F_0}\,{\dot{\psi}}^2\,,
\end{aligned}
\end{equation}
\begin{equation}
\label{equFR11}
\begin{aligned}
&{}-\ddot{\tilde{\sigma}}^2 - 2 H_J \dot{\tilde{\sigma}}^2 - \frac14 \left({\tilde{\sigma}}^2\right)^2  - \left(2 \dot{H}_J + 3 H_J^2\right) \tilde{\sigma}^2 \\
&{}+ \left(6 \dot{H}_J + 12 H_J^2 + \tilde{\sigma}^2\right) \ddot{\tilde{\beta}}_i + \left( 6\ddot{H}_J + 42 H_J \dot{H}_J + 36 H_J^3 + 3 H_J \tilde{\sigma}^2 + \dot{\tilde{\sigma}}^2  \right)\dot{\tilde{\beta}}_i \\
&{} - 3 \left( 2 \dddot{H}_J + 12 H_J \ddot{H}_J + 9 \dot{H}_J^2 + 18 H_J^2 \dot{H}_J \right)=\frac{\varepsilon_\psi}{8 F_0}\,{\dot{\psi}}^2\,, \qquad i = 1, 2, 3.
\end{aligned}
\end{equation}

Combining Eqs.~(\ref{equFR00}) and (\ref{equFR11}) to eliminate $\dot{\psi}$, one can obtain
\begin{equation}
\frac16\left(\ddot{\tilde{\sigma}}^2  + 5 H_J\dot{\tilde{\sigma}}^2  - 2\left(2\dot{H}_J + 3 H_J^2\right)\tilde{\sigma}^2 - \frac{\left({\tilde{\sigma}}^2\right)^2}{2}\right) + \dddot{H}_J + 9 H_J \ddot{H}_J + 3 \dot{H}_J^2 + 18 H_J^2 \dot{H}_J = 0,
\end{equation}
\begin{equation}
\label{sigma_constr}
\dot{\tilde{\sigma}}^2  + 2\tilde{\sigma}^2\, \frac{2\ddot{H}_J + 24 H_J \dot{H}_J + 12 H_J^3 + H_J \tilde{\sigma}^2}{2 \dot{H}_J + 4 H_J^2 + \tilde{\sigma}^2}  = 0.
\end{equation}

In order to resolve these equations and to get one equation with only $\dddot{H}_J$ and another equation with only $\ddot{\tilde{\sigma}}^2$, one needs to differentiate Eq.~(\ref{sigma_constr}) with respect to time. After doing that, one gets the following system of two equations:
\begin{equation}
\label{equD3H}
\dddot{H}_J = \frac{1}{2 \tilde{R}} \left(r_1 - 6r_2 \left[\tilde{\sigma}^2 + 2\dot{H}_J + 4 H^2_J\right]\right),
\end{equation}
\begin{equation}
\label{equD2s2}
\ddot{\tilde{\sigma}}^2 = \frac{3}{\tilde{R}} \left(-r_1 + 4 \tilde{\sigma}^2 r_2\right),
\end{equation}
where
\begin{equation*}
\begin{split}
r_1 &=  \left(\dot{\tilde{\sigma}}^2\right)^2 + \left(4 H_J \tilde{\sigma}^2 + 6 \ddot{H}_J + 36 H_J \dot{H}_J + 24 H_J^3\right) \dot{\tilde{\sigma}}^2\\
 &{} + 2\left(\dot{H}_J \tilde{\sigma}^2 + 14 H_J \ddot{H}_J + 14 \dot{H}_J^2  + 36 H_J^2 \dot{H}_J\right) \tilde{\sigma}^2,
\end{split}
\end{equation*}
and
\begin{equation*}
r_2 = \frac{1}{12}\left(10 H_J \dot{\tilde{\sigma}}^2 - 4\tilde{\sigma}^2\left(2\dot{H}_J + 3 H_J^2\right) - \left({\tilde{\sigma}}^2\right)^2\right) + 9 H_J \ddot{H}_J + 3 \dot{H}_J^2 + 18 H_J^2 \dot{H}_J.
\end{equation*}
Note that an initial condition for $\dot{\tilde{\sigma}}^2$ is connected with other initial conditions by~Eq.~(\ref{sigma_constr}).

The important result is that these equations have a singular point at $\tilde{R} = 0$ if $\tilde{\sigma}^2\neq 0$.
This situation is different from the case of the spatially flat FLRW metric, when $\tilde{\sigma}^2\equiv 0$. Smooth isotropic solutions, with $\tilde{R}$ changing its sign during evolution,
have been found in Ref.~\cite{Ivanov:2021ily}. These solutions have no analogue in the Einstein frame, because the $F(\tilde{R})$ model can be presented in the form of GR model with a standard minimally coupled scalar field only if $F_{,\tilde{R}}=2F_0\tilde{R}>0$.

Using Eq.~(\ref{equFR11}) and relation (\ref{bianchi_constraint}), we get
\begin{equation}
\label{equbeta}
\tilde{R}\ddot{\tilde{\beta}}_i+\left(\dot{\tilde{R}}+3H_J\tilde{R}\right)\dot{\tilde{\beta}}_i=0\,,
\end{equation}
for all $i= 1, 2, 3$.
So,
\begin{equation}\label{tildebetai}
    \dot{\tilde{\beta}}_i=\frac{\tilde{C}_i}{\tilde{a}^3\tilde{R}},\qquad {\tilde{\sigma}}^2=\frac{\tilde{C}_\sigma}{\tilde{a}^6\tilde{R}^2},
\end{equation}
where $\tilde{C}_i$ are constants such that $\tilde{C}_1+\tilde{C}_2+\tilde{C}_3=0$ and $\tilde{C}_\sigma=\tilde{C}_1^2+\tilde{C}_2^2+\tilde{C}_3^2$. For an arbitrary $F(R)$ model, similar results have been obtained in~Ref.~\cite{Gurovich:1979st}. Using relation~(\ref{Rs}), we get for $\tilde{R}_i>0$ that
\begin{equation}
\label{sigma2Ri}
\begin{split}
{\tilde{\sigma}}^2&={}-\frac{2\tilde{R}_i}{3}+\frac{2\tilde{a}^2\tilde{R}_i^2}{3\left[8\tilde{R}_i^3\tilde{a}^{6}+108\tilde{C}_\sigma+12\sqrt{3\left(4\tilde{R}_i^{3}\tilde{a}^{6}\tilde{C}_\sigma+27\tilde{C}_\sigma^2\right)}\right]^{1/3}
}\\&{}+\frac{1}{6\tilde{a}^2}\left[8\tilde{R}_i^3\tilde{a}^{6}+108\tilde{C}_\sigma+12\sqrt{3\left(4\tilde{R}_i^{3}\tilde{a}^{6}\tilde{C}_\sigma+27\tilde{C}_\sigma^2\right)}\right]^{1/3}.
\end{split}
\end{equation}

In this paper, we search for anisotropic solutions, for which  $\tilde{R}$ cannot change its sign during evolution. So, we do not lose smooth solutions if put an additional condition $\tilde{R}>0$. Using this condition, we can get the corresponding Einstein frame model by a conformal metric transformation, find a general solution for this model and get the corresponding solutions for the initial $R^2$ model by an inverse conformal transformation.

\section{Einstein equations in the Bianchi I metric}

If $\tilde{R}>0$, then one can use the Weyl transformation of the metric
\begin{equation}
\label{gmunutrans}
g_{\mu\nu}=\frac{4F_0\tilde{R}}{M^2_{Pl}}\tilde{g}_{\mu\nu},
\end{equation}
and get the chiral cosmological model, described by the following action:
\begin{equation}
\label{SE}
   S_{\mathrm{E}}=\int d^4 x \sqrt{-g} \left[\frac{M_\mathrm{Pl}^2}{2} R -\frac{1}{2}g^{\mu \nu}\nabla_{\mu}\phi \nabla_{\nu}\phi -\frac{\varepsilon_\psi}{2}K(\phi) g^{\mu \nu}\nabla_{\mu}\psi \nabla_{\nu}\psi - \Lambda \right],
\end{equation}
where $M_\mathrm{Pl}$ is the reduced Planck mass,
\begin{equation}
\label{phidF}
\phi=\sqrt{\frac{3}{2}}M_\mathrm{Pl}\ln\left[\frac{4F_0}{M^2_\mathrm{Pl}}\tilde{R}\right]\,,\quad K(\phi)=\mathrm{e}^{\kappa \phi},\quad\kappa={}- \sqrt{\frac{2}{3M^2_\mathrm{Pl}}}\,,\quad\Lambda=\frac{M^4_\mathrm{Pl}}{16F_0}\,.
\end{equation}

The line element for the Bianchi I metric in the Einstein frame is
\begin{equation}
ds^2={}-dt^2+a^2(t)\left[\mathrm{e}^{2\beta_1(t)}\,dx_1^2+\mathrm{e}^{2\beta_2(t)}\,dx_2^2+\mathrm{e}^{2\beta_3(t)}\,dx_3^2\,\right],
\label{Bianchi-I}
\end{equation}
where $a(t)$ is the average scale factor, and $\beta_1(t) +  \beta_2(t) +  \beta_3(t) = 0$.

The shear is
\begin{equation}
 \sigma^2\equiv\dot{\beta}_1^2+\dot{\beta}_2^2+\dot{\beta}_3^2,
\end{equation}
here and hereafter ``dots'' denote derivatives with respect to time $t$.

The evolution equations are:
\begin{equation}
\label{eq_1}
  3H^2-\frac{1}{2}\sigma^2 = \frac{1}{M^2_\mathrm{Pl}} \left(\frac12 \dot{\phi}^2 + \frac{\varepsilon_\psi}{2} K(\phi) \dot{\psi}^2 + \Lambda\right),
\end{equation}
\begin{equation}
\label{trequ} 2\dot H+3H^2+\frac{1}{2}\sigma^2={}-\frac{1}{M^2_\mathrm{Pl}} \left(\frac12 \dot{\phi}^2 + \frac{\varepsilon_\psi}{2} K(\phi) \dot{\psi}^2 - \Lambda\right),
\end{equation}
\begin{equation}
\label{beta_eqn}
  \ddot{\beta}_i = {}-3 H \dot{\beta}_i\,,
\end{equation}
where  $H = \dot{a}/a$\,.

From Eqs.~(\ref{eq_1})---(\ref{beta_eqn}), we get
\begin{equation}
\label{h_avg_eqn}
  \dot{H} + 3 H^2 = \frac{\Lambda}{M^2_\mathrm{Pl}},
\end{equation}
\begin{equation}
\label{equvartheta}
\dot{\sigma}^2={}-6H\sigma^2.
\end{equation}

The field equations are:
\begin{equation}
\label{eq_phi_init}
\ddot{\phi} + 3 H \dot{\phi} - \frac{\varepsilon_\psi}{2} K_{,\phi} \dot{\psi}^2= 0,
\end{equation}
\begin{equation}
\label{eq_psi}
\ddot{\psi}+ 3 H \dot{\psi} + \frac{K_{,\phi}}{K} \dot{\phi} \dot{\psi} = 0.
\end{equation}

Integrating Eq.~(\ref{eq_psi}), one gets
\begin{equation}
\label{dpsi}
    \dot{\psi}=\frac{C_\psi}{a^3K(\phi)}\,,
\end{equation}
where $C_\psi$ is an integration constant.

Subtracting Eq.~(\ref{eq_1}) from Eq.~(\ref{trequ}), we get
\begin{equation}\label{equ21}
2\dot H={}-\sigma^2-\frac{1}{M^2_\mathrm{Pl}} \left(\dot{\phi}^2 + \varepsilon_\psi K(\phi) \dot{\psi}^2\right)={}-\sigma^2-\frac{1}{M^2_\mathrm{Pl}} \left(\dot{\phi}^2 + \frac{C_\psi^2\varepsilon_\psi}{a^6K(\phi)}\right).
\end{equation}
Therefore, in the case of the standard scalar field $\psi$ when $\varepsilon_\psi=1$, $H(t)$ is a monotonically decreasing function and the only solution with a constant $H$ is a de Sitter solution that corresponds to $\sigma^2=0$ and fixed values of the fields.

\section{Solutions in the Einstein frame}

The general solution of this model in the spatially flat FLRW metric has been found in Ref.~\cite{Ivanov:2021ily}.
As one can see Eq.~(\ref{h_avg_eqn}) coincides with the equation for the Hubble parameter in the spatially flat FLRW metric. In the case of a positive $\Lambda$,
Eq.~(\ref{h_avg_eqn}) has the following general solution:
\begin{equation}
\label{H_soln}
  H(t) = \sqrt{\frac{\lambda}{3}}\left[\frac{1 - C \mathrm{e}^{-2\sqrt{3 \lambda} t}}{1 + C \mathrm{e}^{-2\sqrt{3 \lambda} t}}\right],
\end{equation}
where $\lambda \equiv \Lambda / M^2_\mathrm{Pl}$ and $C$ is a constant of integration.

Having an explicit solution for $H(t)$, one can easily integrate Eqs.~(\ref{beta_eqn}) and (\ref{equvartheta}):
\begin{equation}\label{beta_soln}
    \dot{\beta}_i(t)=\frac{C_i \mathrm{e}^{-\sqrt{3\lambda} t}}{1+C\mathrm{e}^{-2\sqrt{3\lambda} t}}\,,
\end{equation}
where $C_i$ are constants of integration.
Therefore,
\begin{equation}\label{sigma2sol}
    \sigma^2(t)=\frac{C_\sigma \mathrm{e}^{-2\sqrt{3\lambda} t}}{\left(1+C\mathrm{e}^{-2\sqrt{3\lambda} t}\right)^2}=\frac{C_\sigma}{4C}\left(1-\frac{3}{\lambda}H^2\right)\,,
\end{equation}
where $C_\sigma=C^2_1+C_2^2+C_3^2=2(C^2_1+C_1C_2+C_2^2)$, since $C_1 + C_2 + C_3 = 0$.

In Table~\ref{T1}, we rewrite this result in the different form, using $t_0 \equiv \ln|C| / (2 \sqrt{3 \lambda})$.

\begin{table}[h]
\begin{center}
\caption{Functions $H(t)$ and $\dot{\beta}_i(t)$ in dependence on value of the integration constant $C$.}
\begin{tabular}{|c|c|c|}
\hline
$C$&   $H\left(t\right)$ & $\dot{\beta}_i \left(t\right)$ \\[1mm]
\hline
$C > 0$
&$\sqrt{\frac{\lambda}{3}} \tanh \left(\sqrt{3 \lambda} (t - t_0)\right)$ &
$\frac{C_i}{\cosh \left(\sqrt{3 \lambda} (t - t_0)\right)}\!$  \\
\hline
$C < 0$
 &  $ \sqrt{\frac{\lambda}{3}} \coth \left(\sqrt{3 \lambda} (t - t_0)\right)$ & $\frac{C_i}{\sinh \left(\sqrt{3 \lambda} (t - t_0)\right)}\!$ \\
\hline
$C = 0$  &  $\sqrt{\frac{\lambda}{3}}$
& $ C_i \mathrm{e}^{-\sqrt{3 \lambda} t}\!$\\
\hline
$C = \pm \infty$ & $-\sqrt{\frac{\lambda}{3}}$ & $ C_i \mathrm{e}^{\sqrt{3 \lambda} t}$ \\
\hline
\end{tabular}
\label{T1}
\end{center}
\end{table}

Using Eq.~(\ref{eq_1}), one can rewrite Eq.~(\ref{eq_phi_init}) in a more convenient form:
\begin{equation}
\label{eq_phi_subst}
\ddot{\phi} + 3 H \dot{\phi} + \kappa \left(M^2_\mathrm{Pl}\left(\lambda - 3 H^2 +\frac{1}{2}\sigma^2\right) + \frac12 \dot{\phi}^2\right) = 0.
\end{equation}

In the case of a nonzero finite $C$, one can use relation (\ref{sigma2sol}) and get Eq.~(\ref{eq_phi_subst}) in the following form:
\begin{equation}
\label{eqn_phi_fin}
\ddot{\phi} + 3 H \dot{\phi} + \kappa \left(M^2_\mathrm{Pl}\left(1 + \frac{C_\sigma}{8 C \lambda}\right)\left(\lambda - 3 H^2 \right)+ \frac12 \dot{\phi}^2\right) = 0.
\end{equation}

By combining Eqs.~(\ref{eq_phi_init}) and~(\ref{eqn_phi_fin}), one gets the following equation:
\begin{equation}
\label{psi_phi_comb}
\varepsilon_\psi K(\phi)\dot{\psi}^2 ={} - 2M^2_\mathrm{Pl}\left(1 + \frac{C_\sigma}{8 C \lambda}\right)\left(\lambda - 3 H^2 \right)- \dot{\phi}^2.
\end{equation}

Introducing $u =\exp(\kappa \phi / 2)$, we linearize  Eq.~(\ref{eqn_phi_fin}):
\begin{equation}\label{equut}
  \ddot{u}+3H\dot{u}-3D\left(3H^2-\lambda\right)\,u=0\,,
\end{equation}
where $D=1/9 + C_\sigma/(72C\lambda)$. It is easy to obtain an asymptotic solution for $u$ from this equation. As $t \rightarrow +\infty$, $3H^2 - \lambda \rightarrow 0$, and one gets
\begin{equation}
\label{utinfty}
u(t) \approx  B_1 + B_2 \mathrm{e}^{-\sqrt{3 \lambda} t},
\end{equation}
where $B_1$ and $B_2$ are constants of integration.

Using $\chi = \sqrt{3 / \lambda}\,H$ as a new variable, we rewrite Eq.~(\ref{equut}) as follows
\begin{equation}
\label{equu}
  \left(1 - \chi^2\right) \frac{d^2 u}{d \chi^2}- \chi \frac{d u}{d \chi} + D u = 0.
\end{equation}

The general solution of this equation is
\begin{equation}
\label{solu}
  u(\chi) = A \cos \left(\sqrt{D}\arccos(\chi) + B\right)=C_1\left(\chi+\sqrt{\chi^2-1}\right)^{\sqrt{D}}+C_2\left(\chi+\sqrt{\chi^2-1}\right)^{-\sqrt{D}},
\end{equation}
where $A$, $B$, $C_1$, and $C_2$ are integration constants. The first expression is useful for $|\chi|\leqslant 1$, whereas the second expression is suitable for $|\chi|\geqslant 1$.
Note that the condition $u(t)>0$ gives restrictions on the integration constants.

It is convenient to write
\begin{equation}
  \phi(t) = {}-\sqrt{6}\, M_\mathrm{Pl}  \ln \left[A \cos \left(\sqrt{D}\arccos\left(\frac{1 - C \mathrm{e}^{-2\sqrt{3 \lambda} t}}{1 + C \mathrm{e}^{-2\sqrt{3 \lambda} t}}\right) + B\right)\right]
\end{equation}
in explicitly real form. We have
\begin{itemize}
\item
for $C > 0$,
\begin{equation}
\label{solphi1}
  \phi(t) = {}-\sqrt{6}\, M_\mathrm{Pl} \ln \left[A \cos \left(\sqrt{D}\arccos\left(\tanh\left(\sqrt{3 \lambda}(t - t_0)\right)\right) + B\right)\right]\,.
\end{equation}
This solution exists at $\varepsilon_\psi=-1$ only, because it corresponds to a monotonically increasing function $H(t)$.

\item
for $C < 0$ and $|C| > C_\sigma / (8 \lambda)$,
\begin{equation}
\label{phi_normal_psi}
  \phi(t) = {}-\sqrt{6}\, M_\mathrm{Pl}\ln\left(C_1 \coth^{\sqrt{D}} \left(\frac{\sqrt{3 \lambda}}{2} (t - t_0)\right)+ C_2 \tanh^{\sqrt{D}}\left(\frac{\sqrt{3 \lambda}}{2} (t - t_0)\right)\right).
\end{equation}

Rewriting the right-hand side of expression (\ref{psi_phi_comb}) in terms of $u$ and $\chi$, and using the rightmost expression in Eq.~(\ref{solu}), we get
\begin{equation}
\varepsilon_\psi  \dot{\psi}^2 = \frac{72M^2_\mathrm{Pl} \lambda\, D }{K^2(\phi)}  C_1 C_2 \left[\coth^2\left(\sqrt{3 \lambda} (t - t_0)\right) -1\right].
\end{equation}
This expression allows us to conclude whether the field $\psi$ has to be phantom or not, depending on the initial conditions. Namely, the only type of solutions for $\phi$ with a non-constant $H$ and a non-phantom field $\psi$ are the ones described by Eq.~(\ref{phi_normal_psi}), and, moreover, the product $C_1 C_2$ has to be positive. If either $C_1=0$, or $C_2=0$, then $\dot\psi=0$.

\item
for $C < 0$ and $|C| < C_\sigma / (8 \lambda)$,
\begin{equation}
\label{solphi3}
  \phi(t) = {}-\sqrt{6}\, M_\mathrm{Pl} \ln \left(A \cos \left[-\frac13 \sqrt{\frac{C_\sigma}{8 |C| \lambda} - 1}\,\mathrm{arcosh} \left(\coth\left(\sqrt{3 \lambda} (t - t_0)\right)\right) + B\right]\right).
\end{equation}
\item
If $C = - C_\sigma /(8 \lambda)$, then Eq.~(\ref{equu}) has the following solution:
\begin{equation}\label{resuu}
    u(\chi)=A+B\ln\left(\chi+\sqrt{\chi^2-1}\right).
\end{equation}
From the condition $u>0$, it follows that $A>0$ and $B>0$.
The corresponding solution $\phi(t)$ is given by
\begin{equation}
\label{solphi4}
\phi(t) ={}-\sqrt{6}\, M_\mathrm{Pl}\ln \left(A \ln \left(\coth \left(\frac{\sqrt{3\lambda}}{2}(t - t_0)\right)\right) + B\right).
\end{equation}
\end{itemize}
Using Eq.~(\ref{psi_phi_comb}), we get that solutions (\ref{solphi3}) and (\ref{solphi4}) corresponds to $\varepsilon_\psi=-1$.
Note that these solutions do not exist in the FLRW metric.

In case of a constant $H=H_0=\pm\sqrt{\lambda/3}$ and $\varepsilon_\psi=-1$, we have the following anisotropic solution:
\begin{equation}
\sigma^2(t)=C_\sigma \,\mathrm{e}^{-6H_0t},\qquad \phi(t) = {}-\sqrt{6}\, M_\mathrm{Pl} \ln \left(A \cos \left(B-\frac{\sqrt{6C_\sigma}}{18 H_0} \,\mathrm{e}^{-3 H_0 t} \right)\right),
\end{equation}
where $A^2=\frac{C_\psi^2}{M_\mathrm{Pl}^2C_\sigma}$.

\section{The connection between the Jordan and Einstein frame solutions of the $R^2$ model}

We have obtained the general solution in the Einstein frame, this solution gives the general solution of the initial $R^2$ model in the parametric time $t$, that is the cosmic time in the Einstein frame.
The metric transformation (\ref{gmunutrans}) connects the metric (\ref{Bianchi-I}) with the following metric in the Jordan frame:
\begin{equation}
ds^2 = {}-\tilde{N}^2(t)dt^2 +\tilde{a}^2(t)\left[\mathrm{e}^{2\beta_1(t)}\,dx_1^2+\mathrm{e}^{2\beta_2(t)}\,dx_2^2+\mathrm{e}^{2\beta_3(t)}\,dx_3^2\,\right],
\label{Fried}
\end{equation}
where
\begin{equation}
\label{Njt}
\tilde{N}(t) =\mathrm{e}^{- \phi(t)/\sqrt{6M^2_\mathrm{Pl}}}, \qquad \tilde{a}(t)=\mathrm{e}^{- \phi(t)/\sqrt{6M^2_\mathrm{Pl}}} a(t).
\end{equation}
So, we get the general solution for the initial $R^2$ model. Note that functions ${\beta}_i(t)$ and $\psi(t)$ are the same in the both frames.

We can find the functions $H_J(\tilde{t})$ and $\tilde{\sigma}^2(\tilde{t})$ in  quadratures.
To get the scalar field $\phi(t)$ that corresponds to the given $H_J(\tilde{t})$ we use
\begin{equation}
\label{Rphi_relation}
6\left(\dot{H}_J(\tilde{t}) + 2 H_J^2(\tilde{t})\right) + \tilde{\sigma}^2(\tilde{t}) = \tilde{R}(\tilde{t}) = \frac{4 \Lambda}{K_0 M^2_\mathrm{Pl}} \mathrm{e}^{\sqrt{2/3}\,\phi(\tilde{t}) /M_\mathrm{Pl}}
\end{equation}
and
\begin{equation*}
t = \int \mathrm{e}^{\phi(\tilde{t})/\sqrt{6 M_\mathrm{Pl}^2}} d\tilde{t} = \int \sqrt{\frac{M^2_\mathrm{Pl}}{4 \Lambda}} \sqrt{\tilde{R}(\tilde{t})}\,d\tilde{t}.
\end{equation*}

The inverse relation is given by
\begin{equation}
\tilde{t}(t) =  \int \mathrm{e}^{-\phi(t) / \sqrt{6 M_\mathrm{Pl}^2}}\, d t=\int u(t)\, d t
\end{equation}
We see that $\tilde{t}(t)$ is a monotonically increasing function. According to Eq.~(\ref{utinfty}), the function $u(t)$ tends to a constant at $t\rightarrow +\infty$. If the function $u(t)$ tends to a nonzero constant at $t\rightarrow +\infty$, then the cosmic time $\tilde{t}$ is proportional to parametric time $t$ at positive temporal infinity. It means that $\tilde{\sigma}^2(\tilde{t})=\sigma^2(\tilde{t}(t))$ tends to zero at $\tilde{t}\rightarrow +\infty$.
Using Eq.~(\ref{solu}), one can see $u(t)$ tends to a nonzero constant for any $C_1\neq -C_2$.  
 At $C_1=-C_2$, it is possible $\tilde{t}(t)$ tends to a finite value $\tilde{t}_{max}$ at $t\rightarrow +\infty$. In this case, $\tilde{\sigma}^2(\tilde{t})$ tends to zero at $\tilde{t}\rightarrow \tilde{t}_{max}$.

For solutions that correspond to $\tilde{R}>0$, we have from Eq.~(\ref{phidF})
\begin{equation*}
\frac{d \phi}{d t}\left(t(\tilde{t})\right) = \frac{d \phi}{d \tilde{t}} \frac{d \tilde{t}}{d t} = \frac{\sqrt{\Lambda}}{\tilde{R}^{3/2}(\tilde{t})}\,\frac{d \tilde{R}(\tilde{t})}{ d \tilde{t}}\,.
\end{equation*}

It is easy to show that
\begin{equation}
H_J(\tilde{t}) = \mathrm{e}^{\phi /\sqrt{6 M_\mathrm{Pl}^2}}\left[H(t(\tilde{t})) - \frac{1}{\sqrt{6} M_\mathrm{Pl}} \frac{d \phi}{d t}\left(t(\tilde{t})\right)\right]
= \frac{1}{u(t(\tilde{t}))} \left[H(t(\tilde{t})) + \frac{\dot{u}(t(\tilde{t}))}{u(t(\tilde{t}))}\right],
\end{equation}
\begin{equation}
\frac{d{\tilde{\beta}}_i(\tilde{t})}{d\tilde{t}} = \mathrm{e}^{\phi(t(\tilde{t}))/\sqrt{6 M_\mathrm{Pl}^2}}\, \frac{d{\beta}_i(t(\tilde{t}))}{dt}\,, \qquad
{\tilde{\sigma}}^2(\tilde{t})=\mathrm{e}^{\sqrt{2/3}\,\phi(t(\tilde{t}))/M_\mathrm{Pl}}\,{\sigma}^2(t(\tilde{t})).
\end{equation}

By using Eq.~(\ref{Rphi_relation}), these equations can also be rewritten as follows:
\begin{equation}\label{HjHe}
 H_J(\tilde{t}) = \frac{2\sqrt{\Lambda\tilde{R}}}{ M_\mathrm{Pl}\sqrt{K_0}}\left[H(t(\tilde{t})) - \frac{1}{2\tilde{R}}\frac{d\tilde{R}(t(\tilde{t}))}{dt}\right]=
 \frac{2\sqrt{\Lambda\tilde{R}}}{ M_\mathrm{Pl}\sqrt{K_0}}\left[H(t(\tilde{t})) - \frac{M_\mathrm{Pl}\sqrt{K_0}}{4\sqrt{\Lambda}\tilde{R}^{3/2}}\frac{d\tilde{R}(t(\tilde{t}))}{d\tilde{t}}\right]\,.
\end{equation}
\begin{equation}
\frac{d{\tilde{\beta}}_i(\tilde{t})}{d\tilde{t}} = \sqrt{\frac{K_0 M_\mathrm{Pl}^2}{4 \Lambda} \tilde{R}(\tilde{t})}\, \frac{d{\beta}_i(t(\tilde{t}))}{dt}\,, \qquad
{\tilde{\sigma}}^2(\tilde{t})=\frac{K_0 M_\mathrm{Pl}^2}{4 \Lambda} \tilde{R}(\tilde{t})\,{\sigma}^2(t(\tilde{t})).
\end{equation}

From Eq.~(\ref{HjHe}), we obtain
\begin{equation}\label{HeHJ}
    H(\tilde{t})=\frac{M_\mathrm{Pl}\sqrt{K_0}}{2\sqrt{\Lambda\tilde{R}}}\left(H_J(\tilde{t})+\frac{1}{2\tilde{R}(\tilde{t})}\frac{d\tilde{R}(\tilde{t})}{d\tilde{t}}\right)\,.
\end{equation}

\section{Conclusion}
$F(R)$ gravity models without scalar fields have anisotropic instabilities associated with the crossing of the hypersurface $ F_{,R}(R)=0$. In other words, the solutions in the FLRW metric are smooth, whereas solutions in the Bianchi I metric have singularities~\cite{Figueiro:2009mm}. It means that the effective gravitational constant cannot change its sign if anisotropy is taken into account. A similar problem has been discussed for the FLRW  and Biachi I models with a nonminimally coupled scalar field~\cite{Starobinsky:1981st} (see also~\cite{Sami:2012uh,Kamenshchik:2016gcy,Kamenshchik:2017ous,Kamenshchik:2017ojc}).

In this paper, we analyze this question for the pure $R^2$ model with a scalar field. While this particular model, in contrast to the Starobinsky inflationary model~\cite{Starobinsky:1980te}, does not include GR as a limiting case, it is a good approximation of the Starobinsky model before inflation and in the beginning of inflation when the $R^2$ term in the action dominates. In the Bianchi I metric, we have found that the evolution equations~(\ref{equD3H}) and (\ref{equD2s2}) have singularity at $\tilde{R}=0$ for anisotropic solutions, whereas for isotropic solutions equations are smooth. So, in distinguish to the case of the spatially flat FLRW metric, for anisotropic solutions, we see that $\tilde{R}$ does not change its sign during evolution.

The $R^2$ gravity model has no ghost if the Ricci scalar $\tilde{R}>0$. We have found anisotropic solutions with $\tilde{R}>0$ using the metric transformation and the Einstein frame. We also analyzed which types of solutions can exist in the case of the phantom scalar field $\chi$ only.
The general solution in  the Einstein frame has been found in terms of elementary functions. This general solution gives explicitly the general solution for the initial $R^2$ model in the parametric time. Solutions in the cosmic time for this model
have been constructed in quadratures.

We plan to generalize our investigation to more complicated $F(R)$ gravity models with the scalar fields and the corresponding two-field chiral cosmological models.

\acknowledgments
V.R.I. is supported by the ``BASIS'' Foundation grant No. 22-2-2-6-1.
\bibliographystyle{apsrev}
\bibliography{BibliographyR2}{}
\end{document}